# R&D on co-working transport schemes in Geant4

**M G Pia**[1], **P Saracco**[1], **M Sudhakar**[1], **A Zoglauer**[2], **M Augelli**[3], **E Gargioni**[4],
[5]**C H Kim, L Quintieri**[6]**, P P de Queiroz Filho**[7]**, D de Souza Santos**[7],
**G Weidenspointner**[9,10]**, M Begalli**[10]
[1]INFN Sezione di Genova, Via Dodecaneso 33, 16146 Genova, Italy
[2]University of California at Berkeley, Berkeley, CA 94720-7450, USA
[3]CNES, 18 Av. Edouard Belin, 31401 Toulouse, France
[4]University Medical Center Hamburg-Eppendorf, D-20246 Hamburg, Germany
[5]Hanyang University, 17 Haengdang-dong, Seongdong-gu, Seoul, 133-791, Korea
[6]INFN Laboratori Nazionali di Frascati, Via Enrico Fermi 40, I-00044 Frascati, Italy
[7]IRD, Av. Salvador Allende, s/n. 22780-160, Rio de Janeiro, RJ, Brazil
[8]MPI für extraterrestrische Physik, Postfach 1603, D-85740 Garching, Germany
[9]MPI Halbleiterlabor, Otto-Hahn-Ring 6, D-81739 München, Germany
[10]UERJ, R. São Francisco Xavier, 524. 20550-013, Rio de Janeiro, RJ, Brazil

E-mail: mariagrazia.pia@ge.infn.it

**Abstract**. A research and development (R&D) project related to the extension of the Geant4 toolkit has been recently launched to address fundamental methods in radiation transport simulation. The project focuses on simulation at different scales in the same experimental environment; this problem requires new methods across the current boundaries of condensed-random-walk and discrete transport schemes. The new developments have been motivated by experimental requirements in various domains, including nanodosimetry, astronomy and detector developments for high energy physics applications.

## 1. Introduction
Geant4 [1],[2] is an object oriented toolkit for the simulation of particle interactions with matter. It provides advanced functionality for all of the the domains typical of detector simulation: geometry and material modelling, description of particle properties, physics processes, tracking, event and run management, user interface and visualisation.

At the present time Geant4 is a mature Monte Carlo system; its multi-disciplinary nature and its wide usage are demonstrated by the fact that its reference article [1] is the most cited publication [3] in the "Nuclear Science and Technology" category of the Journal Citation Reports®.

Geant4 is the result of a research and development project (CERN RD44) carried out between 1994 and 1998. RD44 was launched at a time when the LEP experiments were running GEANT 3 as a well-established system, that had been refined throughout a decade of production service. RD44 investigated the adoption of the object oriented technology and C++ for a simulation system replacing GEANT 3.21 [4], and developed the first functional version of the Geant4 released at the end of 1998.

Since the first release in 1998, new functionality has been added to the toolkit in the following releases; nevertheless, the architectural design and fundamental concepts defining Geant4 application domain have remained substantially unchanged since their original conception.

New experimental requirements have emerged in the recent years, which challenge the conventional scope of major Monte Carlo transport codes like Geant4. Research in nanodosimetry, nanotechnology-based detectors, radiation effects on components in space and at high luminosity colliders, nuclear power, plasma physics etc. have evidenced the need not only of new physics functionality in Geant4, but also of new methodological approaches to radiation transport simulation.

A common requirement has emerged in all such research domains, i.e. the ability to change the scale at which the problem is described and analyzed within a complex environment. This requirement goes beyond the traditional issues of variance reduction, for which current Monte Carlo codes provide abundant tools and techniques.

Significant technological developments – both in software and computing hardware – have also occurred since the mid '90s. New software techniques are available nowadays, that were not yet established at the time when Geant4 was designed.

A research and development project [5] has been recently launched to address fundamental methods in radiation transport simulation to cope with these new experimental requirements and evaluate how they can be supported by Geant4 kernel design. The project focuses on simulation at different scales in the same experimental environment: this set of problems requires new methods across the current boundaries of condensed-random-walk and discrete transport schemes. An exploration is also foreseen about exploiting and extending already existing Geant4 features to apply Monte Carlo and deterministic transport methods in the same simulation environment.

The activity in this research domain was initiated at the Italian Institute of Nuclear Research (INFN); it was approved by the pertinent INFN bodies as a three-year research and development project, named NANO5 in the INFN experiments' database, with formal start in January 2009. The project gathers an international team of physicists and engineers with background in various disciplines: high energy physics, nuclear physics, astronomy and space science and bio-medical physics, as well as software technology. The project also involves the access to experimental data of collaborating groups, which are relevant to the validation of the software.

This paper is associated with a poster related to the NANO5 project presented at the CHEP (Computing in High Energy Physics) 2009 conference; two other contributions related to the NANO5 project were presented in the oral sessions of that conference, and associated papers [6][7] were submitted for publication in the conference proceedings. In order to effectively accommodate the complementary material according to the peculiarities and constraints of the different conference presentation types, the poster, and consequently the related paper, i.e. this one, focused on a general overview of the project, while the details and preliminary results of the first development phase were reported in the oral presentations and related contributions to these proceedings [6][7].

The developments and research results of the NANO5 are intended for publication in scholarly journals. Due to the copyright constraints imposed by the publisher of these conference proceedings, only a subset of the available research material could be included in the present paper and in the complementary ones [6][7] to avoid copyright conflicts with the publishers of the journals where the full account of NANO5 achievements is meant to be published. The publication strategy privileging scholarly journals over conference proceedings is motivated by objective reasons: scientometric analyses have shown the evidence that proceedings have a relatively limited scientific impact compared to scholarly literature in general [8], whereas they play a different role in the scientific community as a vehicle of information about the progress and ongoing activities in the research environment.

More extensive reports, compatible with copyright constraints, of NANO5 research achievements can be found in other conference proceedings [9][10][11][12][13]. The full account of the results of one of the project development sectors has been published in [14], while the proceedings of the CHEP

2009 conference were under review. Other papers related to NANO5 are currently in preparation in view of their publication in scholarly journals.

## 2. Main areas of research and development

The research and development addresses two main areas in radiation transport, which are both characterized by a common conceptual intent: the capability of dynamically adapting the simulation to different transport schemes according to the environment. They concern the simulation at different scale, which would require new methods across the current boundaries of condensed-random-walk and discrete methods, and the possibility of exploiting Monte Carlo and deterministic transport methods in the same simulation environment in cases.

Addressing such issues implies developing conceptual methods and innovative design solutions in a Monte Carlo kernel system. The Geant4 toolkit is the ideal playground for this research and development, thanks to the object oriented technology it adopted in the RD44 phase.

Other issues have been identified along with the experience of Geant4 development and usage over the past 10 years, which would profit from research and development in the kernel design:
- Customization of physics modeling in a simulation application
- Scattered and tangled concerns across the code
- Facilities for physics verification and validation
- Performance

These topics are considered as supporting developments, which are instrumental to achieve the main goals of the project.

## 3. Co-working condensed and discrete simulation methods

One of the research activities in the inception phase of the NANO5 project concerned the collection of the experimental requirements concerning the treatment of particle transport according to condensed and discrete simulation methods.

Methods to model hard interactions of particles with matter constituents by means of an appropriate binary theory are well established: in this approach collisions are treated as binary processes, that is, either the target electrons are treated as free and at rest, or the influence of binding is accounted only in an approximated way.

General-purpose Monte Carlo codes, like EGS [15]-[17], FLUKA [18]-[19], Geant4 and MCNP [20]-[22], operate in this context. Their calculations of energy deposit distributions are based on condensed-random-walk schemes of particle transport. Charged particle tracks are divided into many steps, such that several interactions occur in a step; one energy loss and one deflection are calculated for each step. A further simplification consists in the adoption of the Continuous Slowing Down Approximation (CSDA), where the energy loss rate is determined by the stopping power. This approach is adequate as long as the discrete energy loss events treated are of magnitudes larger than electronic binding energies.

Various specialized Monte Carlo codes, usually known as "track structure codes", have been developed for micro/nano-dosimetry calculations. They handle particle interactions with matter as discrete processes: all collisions are explicitly simulated as single-scattering interactions. This approach is suitable to studies where the precise structure of the energy deposit and of the secondary particle production associated with a track is essential; nevertheless, the detailed treatment of collisions down to very low energy results in a high computational demand.

So far, simulation based on condensed-random-walk schemes and track structure generation have been treated as separate computational domains. This separation is due to the conceptual and technical difficulty of handling the two schemes in the same simulation environment. Achieving a conceptual approach and an architectural design where the two schemes can co-work would represent fundamental progress in Monte Carlo simulation.

Recently, a set of specialized processes for track structure simulation in liquid water has been designed and implemented in Geant4 [23]; like their equivalents in dedicated Monte Carlo codes, they operate in the régime of discrete interactions. While the toolkit nature of Geant4 allows the co-existence of tools for simulation at different scales, the capability of these two schemes to effectively co-work in a multi-scale problem is still far from being established.

The issue of co-existing condensed-random-walk and discrete schemes arises in another context of the simulation domain. It concerns the conceptually correct treatment of the atomic relaxation following the impact ionization produced by charged particles: since the cross section for producing secondary electrons from ionization is subject to infrared divergence, in the conventional condensed-random-walk schemes the interaction is treated in two different régimes of continuous energy loss along the step and of discrete δ-ray production, with the consequent adoption of cuts. This scheme introduces an artificial dependency on cuts in the generation of PIXE (Particle Induced X-ray Emission), while atomic relaxation is intrinsically a discrete process. Moreover, the current Geant4 scheme neglects the correlation between the δ-ray spectrum of primary ionization and PIXE. Therefore a conceptual revision of the continuous energy loss and discrete scheme is desirable in this physics domain too.

The achievement of co-working transport schemes is relevant to various experimental domains. The analysis of the impact that evolutions in transport methods could have on experimental capabilities was performed in the context of the Business Modeling discipline of the Unified Process [24]. The results of this activity provide guidance for the software development in the following phases of the project and contribute to shape the various branches of the project compatible with priorities in the multi-disciplinary scientific environment of its software applications.

Radiation effects at the nano-scale are important for the protection of electronic devices operating in various radiation environments. In realistic use cases such small-scale systems are often embedded in larger scale ones: for instance, a component may operate within a HEP experiment or on a satellite in space, cellular and sub-cellular aggregates in real biological systems exist in complex body structures etc. Applications of a simulation system capable of addressing different scales would be relevant to studies of the effects on components exposed to the fierce experimental environment of LHC (and super-LHC).

The capability of transition across condensed and discrete Monte Carlo simulation schemes in the same software environment is also critical to experimental configurations involving nanotechnology-based detectors. While research and development in nanotechnology is actively pursued also in view of application to HEP detectors, it is not yet possible to simulate such detectors as standalone systems with Geant4, nor to evaluate their performance once they are embedded in a full-scale experimental set-up.

Plasma physics requires addressing the concept of object state and behaviour mutation in relation to the environment: in this use case the mutability concerns both the physics processes and the particles involved. Astrophysics and studies for fusion-based nuclear power are just two relevant applications, which would profit from Geant4 applicability to this physics domain.

Use cases affected by the current conceptual limitation to treat PIXE correctly in Geant4 involve multiple, multi-disciplinary domains, including applications for material analysis, planetology, cultural heritage, precise dosimetry and critical shielding optimization of X-ray telescopes.

## 4. Co-working Monte Carlo and deterministic simulation methods

Deterministic transport methods are widely used in various domains: radiotherapy treatment planning and calculations for nuclear reactors are just two examples. Their usage is motivated by the requirement of a fast simulation response in complex situations.

Both Monte Carlo and deterministic transport methods have their own strong points and limitations. A research and development in this field to perform both transport models within the same simulation environment would be highly valuable: the capability of different transport methods in the same simulation environment would simplify the geometrical and material modelling of the system under study, and would facilitate the analysis of the behaviour of the system itself.

Similarly to what discussed in the previous section, the application of Business Modelling techniques to the analysis of this sector of the problem domain has identified objective criteria to guide the following phases of the software development process.

**5. Supporting research and development topics**

The complexity of the problem domain to be addressed requires the investigation of software techniques, capable of effectively supporting the conceptual objectives to be pursued.

5.1. Generic programming technology in physics simulation design

Metaprogramming has emerged in the last few years as a powerful design technique. In C++ the template mechanism provides naturally a rich facility for metaprogramming; libraries like Boost and Loki are nowadays available to support generic programming development. Metaprogramming presents several interesting advantages, which make it as a worthy candidate for physics simulation design.

This technique has not been exploited in Geant4 core yet: the evolution towards the C++ standard still in progress and the limited support available in C++ compilers in the mid 90's prevented the exploitation of templates in Geant4 architectural design during the RD44 phase. A first investigation of its applicability in a multi-platform simulation context has been carried out by one of the authors of this paper through the application of a policy-based class design [23] limited to a small physics sub-domain.

A broader evaluation of the suitability of this technique for application to physics simulation is addressed in the context of the NANO5 project, also involving the acquisition of quantitative metrics. Preliminary results in this area of the project are reported in [6].

5.2. Design for scattered and tangled concerns

The problem domain of radiation transport simulation involves a number of concerns, which are common to multiple parts of the system, but whose code gets scattered across different parts; moreover, multiple concerns may be tangled in the same code. The capability of addressing scattered concerns by an effective design would results in leaner, more easily maintainable Monte Carlo code: this is a non-negligible issue for optimally exploiting the available resources in a large-scale software system like Geant4.

Two topics associated with cross-cutting concerns in Geant4 code are relevant to the research areas considered in this project: the issue of endowing objects - in particular, physics objects - of intrinsic verification and validation capabilities (more in general, of analysis capabilities), and dealing with secondary effects following a primary interaction (e.g. the relaxation of an excited atom).

The object oriented technology lacks proper instruments to address the issue of scattering and tangling of concerns. Aspect oriented programming (AOP) provides support for cross-cutting concerns (i.e. aspects) and for automatically propagating appropriate points of execution in the code; nevertheless, this technology is not widely established yet, and language support is still relatively limited in C++. Therefore a feasibility study would be useful to evaluate how to address this issue in the design of a large scale simulation system written in C++.

At the present stage of the project, the activity in this area regards the identification and analysis of scattered and tangled concerns throughout Geant4 electromagnetic physics code.

5.3. Native testing and analysis of physics simulation software

The Verification and Validation [25] process plays a fundamental role to ensure the quality of the software. A simulation system is especially concerned, due to the critical role it plays in the experimental lifecycle and sensitive applications like medical physics, radiation protection etc.

A software design enabling intrinsic Verification and Validation capabilities in the physics code itself would greatly facilitate this crucial process by increasing the robustness of the software and reducing the need of dedicated resources. Therefore the possibility of endowing the software with intrinsic testing capability is investigated in the context of this project; should this first research and development demonstrate a successful solution, further extension to other Geant4 domains could be envisaged.

Preliminary achievements in this area are described in [6], [10] and [11]; they show how some critical choices in the physics design greatly facilitate the software verification process and its validation against experimental data, and provide a first quantitative evaluation of achievable improvements.

## 6. First developments

At the time of the conference associated with these proceedings the software development process of the project was in the inception phase, according to the dynamic dimension of the adopted software process model, based on the Unified Process [24]. Particular emphasis was invested in this phase in the discipline of requirements and in the problem domain analysis; pilot projects, involving the development of exploratory prototypes, support the clarification of the requirements, the domain analysis and the investigation of candidate design solution.

The issue of intertwined condensed and discrete physics processes was studied in the context of a pilot project dedicated to PIXE (Particle Induced X-ray Emission). Preliminary results are reported in [7] and [12]; an extensive report has been published in [14], which describes the new developments in detail and demonstrates their application to a first concrete experimental use case.

The issue of decomposing physics processes into stable and mutable parts has been addressed through two pilot projects devoted to the investigation of generic programming techniques in support of this requirement, respectively associated with processes originally conceived for conventional condensed transport schemes and for track structure simulation.

One of these pilot projects concerns the redesign and implementation of the current Geant4 photon interaction processes in a policy-based class design approach; preliminary results are documented in [6], [10] and [11], concerning the impact of the candidate design on various aspects like functionality, computational performance and ease of testing.

The other pilot project involves reengineering the functionality of a FORTRAN code originally developed for nanodosimetry simulation [26] in a design compatible with Geant4 kernel; the initial activity in this context has evaluated the suitability of the policy-based design developed for photon processes to support discrete transport models for electron interactions as well. The first results of this pilot project are documented in [13]; apart from the original intended purpose of evaluating prototype design solutions, they contribute also an extension of Geant4 physics functionality.

## 7. Conclusion and outlook

A research and development project is in progress to address the capability of handling multi-scale use cases in the same simulation environment associated with Geant4: this requirement involves the capability of handling physics processes according to different transport schemes. Research and development is also in progress to evaluate design techniques, like generic programming and handling concerns, capable of supporting the main design goals of the project.

The activity of the project, in the inception phase at the time of the CHEP 2009 conference, has been initially focused on the disciplines of Requirements and Business Modeling to identify the needs of the experimental community in multidisciplinary applications and the impact of new particle transport capabilities. A set of prototype developments has investigated candidate design solutions to address some key issues in the problem domain.

Further details concerning the first developments associated with this project are described in two other contributions [6][7] to these conference proceedings; preliminary results have been documented in various references concerning exploratory investigations associated with different, complementary aspects of the project.

Detailed reports of the prototype developments and related results briefly outlined in this paper are included in journal publications currently in preparation.

**Acknowledgments**
The authors thank S. Bertolucci (CERN), T. Evans (ORNL), S. Giani (CERN), B. Grosswendt (retired, former PTB), A. Montanari (INFN Bologna), A. Pfeiffer (CERN), R. Schulte and A. Wroe (Loma Linda Univ.) for helpful discussions and support to the project.